\documentclass[sigconf]{acmart}

\usepackage{hyperref}
\usepackage{url}
\usepackage{color}

\AtBeginDocument{%
  \providecommand\BibTeX{{%
    \normalfont B\kern-0.5em{\scshape i\kern-0.25em b}\kern-0.8em\TeX}}}

\renewcommand\footnotetextcopyrightpermission[1]{} % removes footnote with conference information in first column
\begin{document}

\title{Towards a Backdoorless Network Architecture Based on Remote Attestation and Backdoor Inspection}


\author{Takayuki Sasaki, Yusuke Shimada}
\affiliation{\institution{NEC Corporation}}
\email{{tsasaki,shimada_2019}@nec.com}
%\author{Anonymous authors}

%\renewcommand{\shortauthors}{Sasaki and Shimada}

\begin{abstract}
To keep a system secure, all devices in the system need to be benign. To avoid malicious and/or compromised devices, network access control such as authentication using a credential and remote attestation based on trusted hardware has been used. These techniques ensure the authenticity and integrity of the devices, but do not mitigate risks of a backdoor embedded in the devices by the developer. To tackle this problem, we propose a novel architecture that integrates remote attestation and backdoor inspection. Specifically, the backdoor inspection result is stored in a server and the verifier retrieves and checks the backdoor inspection result when the remote attestation is performed. Moreover, we discuss issues to deploy the proposed architecture to the real world.
\end{abstract}





\maketitle
\pagestyle{plain} % removes running headers

\section{Introduction}

Recently, network systems become complex and the systems are constructed using multi-vendor devices. Moreover, a device may comprise many components, and part of the components may come from external suppliers. In such a situation, we can identify supply-chain risks, where malicious devices exist in the system. There are some types of supply-chain risks, for example, information leakage from subcontractors or insufficient quality of the software. Particularly, in this paper, we focus on the risks of backdoors which are embedded by malicious device vendors or malicious insiders in the device vendor.

We can find news about actual backdoors, for example, devices with backdoor such as a router~\cite{huawei}, IP camera~\cite{ipcamera}, and laptop\cite{laptop} have been reported. Moreover, it is also reported that backdoors are embedded in the software development phase by tampering with C/C++ libraries~\cite{tmreport}.

To avoid the connection of malicious devices, there are some existing techniques. Network access control using a credential, e.g. IEEE802.1X, performs authentication when a device is connected with the network. Trusted Network Connect~(TNC)~\cite{tnc} performs remote attestation based on a Trusted Platform Module~(TPM) and allows only the non-tempered device to connect to the network. Specifically, TNC measures the integrity of the remote device and obtains the hash values of the device, then TNC compares the retrieved hash value with the correct hash value in the whitelist.

Unfortunately, TNC only ensures the integrity of the device and backdoor risks embedded by the device vendor cannot be mitigated. Specifically, in case that the backdoor is already in the device when the correct hash value is calculated during the manufacturing process, the remote attestation using the hash value cannot detect the backdoor, because the hash value at device manufacturing and the hash value calculated at the remote attestation become the same.

To cope with the problem, we propose an architecture that integrates backdoor inspection and the remote attestation. At the backdoor inspection, the backdoor inspection certificate is issued and registered to a server. Then, when performing the remote attestation, the verifier retrieves the backdoor inspection certificate from the server. The verifier accepts the device connection only when the device is not tampered with and the backdoor inspection certificate shows that no backdoors are found.
Even using this architecture, we cannot ensure any backdoor in the device due to false positives of backdoor detection algorithms, but at least we can ensure the backdoor inspection is performed and the risks of the backdoor are mitigated.

In this paper, we further discuss issues when the architecture is deployed. For example, to enhance security, the proposed architecture needs to be integrated with existing security technologies such as monitoring and supply chain management. Moreover, we discuss the concrete operations of the architecture.

In summary, our major contributions are as follows.
\begin{itemize}
    \item Network access control architecture to mitigate backdoor risks by integrating backdoor inspection and remote attestation
    \item Discussion of issues when the architecture is deployed to the real world
\end{itemize}

\section{Background and problem statement}
\subsection{Scope of the backdoor}
There are two types of backdoor: hardware-based and software-based.
As for the hardware-based backdoor, the malicious functions are implemented in the hardware chips. Thus, analysis of the hardware-based backdoor is difficult, but some techniques have been proposed~\cite{hw-trojan}. As for the software-based backdoor, we can apply software analysis techniques for detecting the backdoor. In this paper, we focus on the software-based backdoor.

Software-based backdoors can be embedded mainly at two different phases: backdoor embedded at the development phase and backdoor provided by malware which is infected at the running phase of the device.
In this paper, we focus on the backdoor at the development phase because the backdoor created by the malware can be detected using the integrity measurement of the remote attestation and other malware detection techniques.

\subsection{Backdoor detection techniques}
Backdoor detection is difficult compared to malware detection because it is hard to distinguish legitimate functions and backdoor functions. There are a few approaches to detect backdoors. For example, control flow analysis for detecting authentication bypass~\cite{firmalice}, classification of binaries for detecting hidden functionalities~\cite{humidify}, and detection and scoring of static data comparison~(e.g. fixed string) for detecting hidden credentials~(e.g. hidden admin account)~\cite{stringer}.

There is no perfect algorithm for backdoor detection, and each algorithm has its advantages and disadvantages. Thus, to detect various types of backdoors, inspection using multiple algorithms is required. Moreover, these algorithms cause false negatives, thus we cannot guarantee no-backdoor for sure. In addition, false positives could be included in the detection results. Furthermore, the attacker would evade the backdoor detection. For example, binary obfuscation techniques embed dummy instructions and/or dummy control flows, and also the firmware could be encrypted.

The backdoor detection algorithms themselves are beyond the scope of this paper and we utilize existing algorithms. We introduce the algorithms in Section~\ref{sec:backdoor-detection}.

\subsection{Remote attestation}
To check the integrity of a remote entity, remote attestation based on trusted boot using TPM has been proposed. In the trusted boot procedure, TPM measures BIOS, BIOS measures OS, and OS measures applications. By this boot procedure, the trust chain is constructed from TPM as a root of trust to applications via BIOS and OS.

In remote attestation procedure, TPM outputs a report about the measurement result of the entire system and adds a signature to the report. The verifier can confirm the integrity of the remote entity by checking the report and its signature. Specifically, the hash value in the report and the correct hash value which is calculated in advance are compared.
TNC allows the connection of a device, only when the remote attestation shows that the device is not tampered.

\subsection{Problem statement}
Remote attestation and TNC based on the remote attestation are insufficient to mitigate backdoor risks. This is because the remote attestation requires the correct hash value as a clean state of the device, but in case that the backdoor is already embedded before the calculation of the correct hash value, the remote attestation assumes the device with the backdoor as the correct state.

The problem is that there is no association between the backdoor inspection and remote attestation. Thus, the goal of this paper is to propose an architecture that integrates the backdoor inspection and the remote attestation. Specifically, the architecture refuses devices for which backdoor inspection is not performed and also refuses devices where backdoors are detected by the inspection.

\subsection{Assumptions of security model}
In this paper, we assume the following items for the backdoor developer and backdoor inspector.
\begin{itemize}
    \item Backdoor is embedded as a part of software by a malicious developer or a malicious device manufacturer at the development phase of a device.
    \item The backdoor inspector can access the firmware of the device. 
\end{itemize}
\section{Architecture}
Here, we propose the architecture to tackle the problem discussed in the previous section.

\subsection{Overview}
To integrate the backdoor inspection and remote attestation, in the proposed architecture, the calculation of the hash value used for the remote attestation is done during the backdoor inspection. Then, the hash value is recorded in the backdoor inspection report.
At the remote attestation, the verifier can leverage the backdoor inspection result by comparing the hash value retrieved by the remote attestation and the hash value recorded in the backdoor inspection certificate.

\begin{figure}[t]
  \begin{center}
    \includegraphics[width=7.0cm]{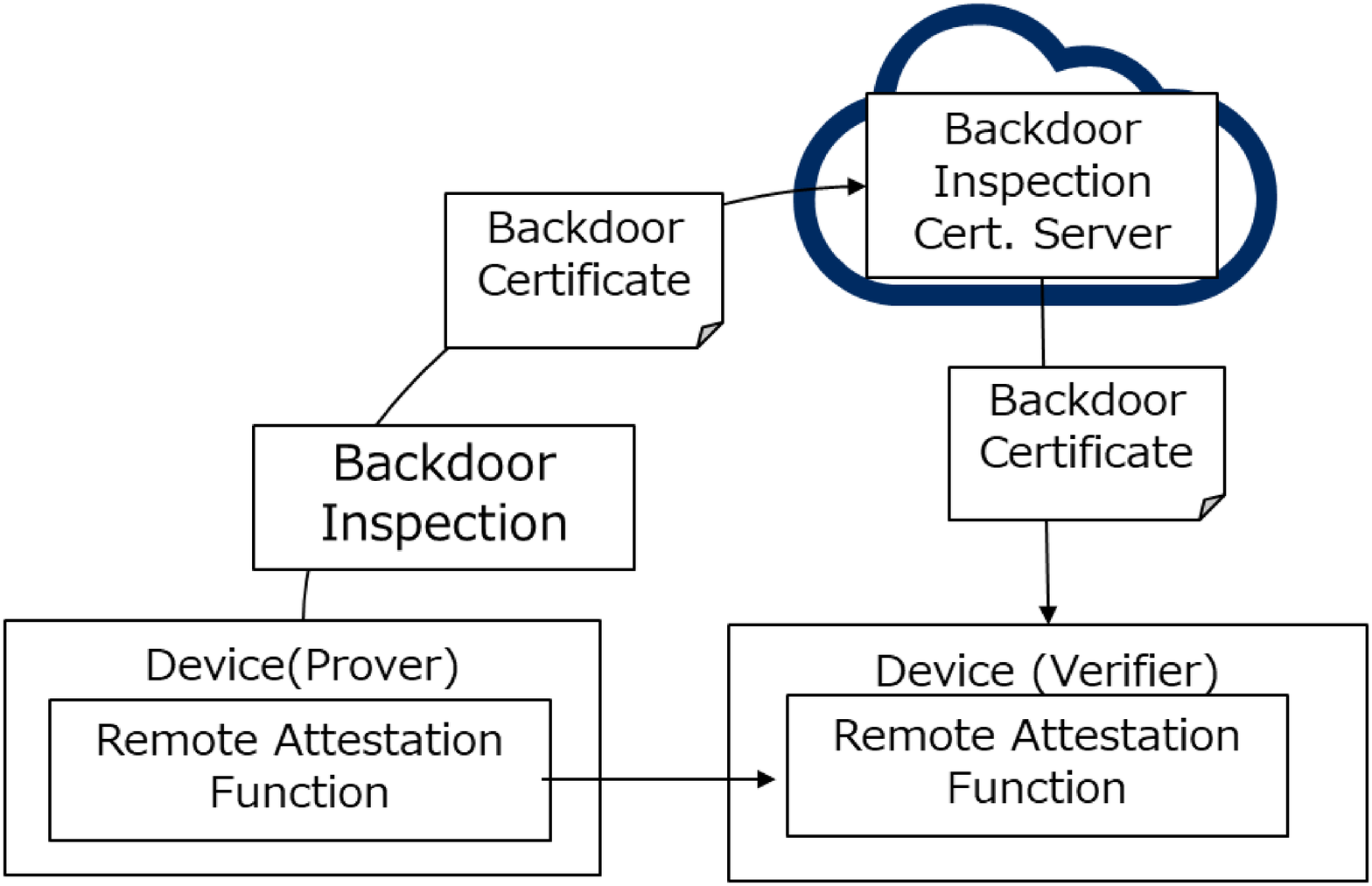}
    \caption{Architecture}
    \label{fig:architecture}
  \end{center}
\end{figure}

Figure~\ref{fig:architecture} shows the proposed architecture. In the proposed architecture, there are two major steps. The first step is for backdoor inspection and issuing its certificate. The second step is for network access control using the backdoor inspection certificate. In the following, we describe the details of the steps.

\subsection{Backdoor inspection}
\label{sec:architecture-backdoor-inspection}
Backdoor inspection is performed using the existing algorithms. In this step, a plurality of backdoor detection algorithms is used to expand the coverage of the backdoor types.

In addition to the backdoor inspection, general vulnerability detection can be also performed to minimize the risks of vulnerability-based backdoor~\cite{ufo}.

Based on the backdoor inspection, a backdoor inspection certificate is issued. The certificate can contain the following information.
\begin{itemize}
    \item Backdoor inspection result. A list of backdoor detection algorithms and these results are recorded. In addition, the parameters of the algorithm are also recorded for checking the algorithms are executed with correct configurations. In some algorithms, the detection results are output as scores rather than black and white decision. In this case, the scores are recorded as the results.
    \item Hash value of the software. As described above, to associate the backdoor inspection result and the software state, the hash value of the software is recorded.
    \item Name of the organization that performs the backdoor inspection and its signature. To clarify the inspector organization, its name is recorded. Moreover, to avoid modification of the backdoor inspection certificate, the organization adds a signature to the certificate.
    \item Name of the backdoor inspection engineer or evidence that shows the inspector has qualifications/skills to find the backdoor. Some backdoor algorithms require manual operations and/or final decision by the backdoor inspection engineer. In such a case, the confidence of the certificate depends on the skill of the engineer. Thus, as extra information, the name of the engineer or qualification/skills can be recorded.
\end{itemize}

Finally, the backdoor inspection certificate is uploaded to the backdoor inspection certificate server by the backdoor inspector.

\subsection{Network access control based on the backdoor inspection}
\label{sec:netowork-access-control}
\begin{figure}[t]
  \begin{center}
    \includegraphics[width=9.0cm]{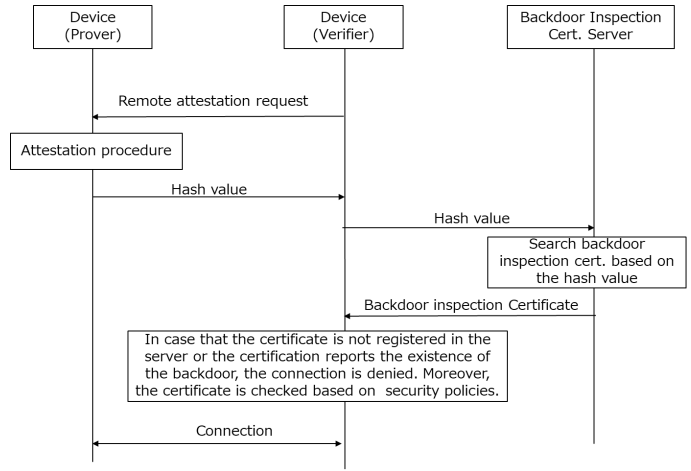}
    \caption{Remote attestation sequence}
    \label{fig:sequence}
  \end{center}
\end{figure}

The network access control is performed on the basis of the remote attestation result, the backdoor inspection certificate, and pre-defined security policies~(Figure~\ref{fig:sequence}). Here, the prover is a device that wants to connect to the system and the verifier is a device for network control such as a switch, a firewall, etc.
First, the remote attestation is performed and the verifier obtains the hash value from the prover. Then, the verifier sends the hash value to the backdoor inspection certificate server. The server searches the certificate using the hash value as a key and returns the certificate to the verifier. Finally, the verifier makes a decision on the basis of the certificate and security policies.

Here, the security policies can include the following configuration items.
\begin{itemize}
    \item Inspected backdoor types. To specify a requirement for the coverage of the backdoor inspection, inspected backdoor types are designated. The verifier compares this item and one in the backdoor inspection certificate.
    \item Backdoor detection algorithms. To check whether the backdoor inspection is performed using correct methods, this item is specified. The verifier checks backdoor detection algorithms in the certificate based on this item.
    \item Backdoor inspection organization. To ignore backdoor inspection certificates from untrusted/malicious inspector organizations, the verifier specifies the organizations that can be trusted. Only certificates from the specified organizations are valid.
    \item Obligations after allowing the connection from the device. The backdoor detection results would be grey. In this case, the connection of the device is allowed under extra security measures such as monitoring. We discuss it in Section~\ref{sec:other-security-measures}.
\end{itemize}

Connections of the devices that do not meet the security policies are rejected in addition to the devices without the backdoor inspection certificate and devices in which the certificate mentions the existence of the backdoor.

\section{Discussion}
The proposed architecture is simple and could be implemented, but there would be issues for deployment of the architecture in the real world. Here, we discuss the issues.

\subsection{Combination of other security measures}
\label{sec:other-security-measures}
The existing backdoor detection algorithms are not perfect and would miss the backdoors~(false negatives). Moreover, the algorithm would output the 'grey' result and require additional manual analysis. However, the manual analysis requires cost, time, and trained engineers.
Thus, for low cost backdoor inspection, the inspection only by the automated algorithm can be considered.
In such cases, to mitigate risks of missed backdoors and  grey devices, the following extra security measures~(obligations) can be enforced to the grey devices.

\paragraph{Run-time monitoring.}
Intrusion detection system~(IDS) which detects abnormal behaviors using machine learning techniques can be performed as an extra security measure. In this case, the threshold of the abnormal behaviors algorithm can be modified based on the backdoor inspection scores.Specifically, a more strict threshold is applied to the more suspicious devices.
Moreover, detailed logging can be performed on grey devices to efficiently detect malicious behaviors.

\paragraph{Access control.}
In addition to the control of the connectivity discussed in Section~\ref{sec:netowork-access-control}, fine-grained access control can be performed. For example, if one software component is grey, we can enforce system call-level access control such as SELinux so that the software component can only have minimal permissions for its functions. In addition, internal network access control such as IP level filtering and VLAN level isolation can be enforced for the grey devices in order to minimize the damage in case that the devices have the backdoor.

\subsection{Integration with supply chain management}
In addition to the above technical measures, an ecosystem of a secure supply chain can be enforced as well. For example, background checks and internal control/corporate governance checks can be performed for the suppliers. 

This supply chain information also can be recorded in the backdoor inspection certificate and used for the decision of the device connection. For example, using this information a verifier can check that a device is constructed using components of trusted suppliers. 

\subsection{Who performs the backdoor inspection?}
\label{sec:inspector}
We can find a question: who should inspect the backdoor? For its answer, we can identify the following three cases.
\begin{itemize}
    \item Trusted third parties. For example, the government or a public organization inspects devices. This option is the most reliable but there would be a scalability issue that an organization cannot inspect all devices.
    \item Organizations using devices. The users of the device inspect the devices. In this case, an issue is that there would be no backdoor inspection engineers, especially for small and medium-sized businesses.
    \item Device manufacturers. Assuming that the device manufacturers are malicious, this option does not work. However, in a case that the device manufacturers are benign and their suppliers are potentially malicious,  backdoor inspection by the device manufacturers is possible.
\end{itemize}

\subsection{Disclosure of backdoor detection algorithm}
To trust the backdoor inspection result in the certificate, the verifier would like to know the backdoor inspection procedures including how to find the backdoor. However, in case that the details of the backdoor detection algorithms are disclosed, the backdoor developers may try to bypass the detection algorithm.

To avoid the cat and mouse game, a fact that shows the backdoor inspection is performed in specified procedures could be only disclosed. At least, consensus between backdoor inspectors and device users~(verifiers) is required.

\subsection{Software update}
The software of the devices are often updated for new functionalities and fixing vulnerabilities. In the case of the update, the backdoor inspection should be performed again. The backdoor inspection is only performed for the updated parts of the software and the backdoor inspection certificate is also updated.

\subsection{Application to secure system construction}
In the above, we discuss the network access control when a device connects to a system. In addition to this use case, our architecture is useful for system construction. For example, when a system integrator would like to internally use a device from third parties, the system integrator can leverage the backdoor certificate of the device.
\section{Related work}
Here, we introduce backdoor detection techniques.

%\subsection{Backdoor detection techniques}
\label{sec:backdoor-detection}
Some researchers propose methods to detect software backdoors in semi-automated or automated fashion.
Schuster et al.~\cite{Schuster} introduce an approach to automatically detect and disable certain types of backdoors within server applications. They use dynamic analysis technique to automatically identify the specific regions in a binary that are prone to attacks, such as authentication routines or command dispatching or handling functionalities. They then leverage this knowledge to determine suspicious components in a semi-automated manner.
Firmalice~\cite{firmalice} is designed for detecting authentication bypass vulnerabilities within embedded device firmware.
The authors firstly define the security policy to denote privileged operations in the program, and then utilize symbolic execution techniques to detect whether it is possible to reach the point of executing the privileged operations without proper authentications.
Thomas et al. propose two different methods to detect backdoor components, HumIDIFy~\cite{humidify} and Stringer~\cite{stringer}.
HumIDIFy uses a machine learning technique to classify binaries from Linux-based embedded device firmware into  functionality classes such as web-server or secure-shell daemon. Then, it evaluates the binary based on a class-specific profile and automatically detects the deviations from their expected functionalities.
Stringer aims to identify comparisons with static data, which guards the succeeding unique functionality.
It covers not only hard-coded credential checks, but also undocumented functionalities within the targeted binary. 

The firmware of devices is sometimes legitimately obfuscated or packed/encrypted to protect intellectual properties.
To analyze such firmware, we can integrate existing deobfuscation techniques with the above backdoor detection techniques. 

As discussed in Section~\ref{sec:architecture-backdoor-inspection}, these algorithms can be integrated with the proposed architecture.
\section{Conclusion and future work}
Remote attestation is insufficient to mitigate risks of backdoor embedded in the devices by the device manufacturers. To tackle this problem, we propose the architecture where the decision of a network connection acceptance is made on the basis of the remote attestation result, the backdoor inspection certificate, and the security policies. We further discuss deployment and operation issues.

As for future work, we will implement the architecture and evaluate its feasibility. In addition, we will discuss non-technical frameworks such as consensus on the backdoor inspection procedure and organizational rules.
Moreover, the reliability of the proposed architecture depends on the accuracies of the backdoor detection algorithms, thus we will investigate these accuracies and improve the detection capabilities.

The proposed architecture is still a work in progress, but we believe that the architecture contributes to improving the trustworthiness of systems.

%\section*{Acknowledgments}
%The author would like to thank \textbf{Anon.} for proof-reading and valuable feedback.

\bibliographystyle{acm}
\bibliography{reference}

\end{document}